# ptp++: A Precision Time Protocol Simulation Model for OMNeT++ / INET

Martin Lévesque and David Tipper

School of Information Sciences
University of Pittsburgh
Pittsburgh, PA, USA

*Abstract*—**Precise time synchronization is expected to play a key role in emerging distributed and real-time applications such as the smart grid and Internet of Things (IoT) based applications. The Precision Time Protocol (PTP) is currently viewed as one of the main synchronization solutions over a packet-switched network, which supports microsecond synchronization accuracy. In this paper, we present a PTP simulation model for OMNeT++ INET, which allows to investigate the synchronization accuracy under different network configurations and conditions. To show some illustrative simulation results using the developed module, we investigate on the network load fluctuations and their impacts on the PTP performance by considering a network with class-based quality-of-service (QoS) support. The simulation results show that the network load significantly affects the network delay symmetry, and investigate a new technique called class probing to improve the PTP accuracy and mitigate the load fluctuation effects.**

## I. INTRODUCTION

Precise time synchronization is a key requirement in several packet based communication networks and real-time networked application domains, such as the automated industrial systems and smart power grid systems. For instance, the communication technologies such as the Long Term Evolution-Advanced (LTE-A) cellular networks require backhaul equipment base stations to provide time synchronization in order to synchronize transmissions over frequencies from adjacent base stations and interference coordination. Also, the emerging smart power grid systems, which are characterized by a two-way flow of energy and end-to-end communications, will also require tight time synchronization. In those systems, the communications are machine-to-machine in nature and require synchronized information in order to improve reliability and efficiency of power delivery.

A general solution to provide the synchronization functionality among networked devices requiring time alignment is to incorporate an atomic clock or a Global Positioning System (GPS) component in each device. However, equipping each device of these technologies would be extremely costly, especially for instance in sensor and actuator devices, which in many applications are cost and computationally constrained.

To reduce these cost issues, network protocols have been developed to distribute time over packet-switched networks and synchronized distributed devices. The Network Time Protocol (NTP)[1] is a widely used Internet time synchronization protocol, which provides millisecond synchronization accuracy and is implemented at the application layer. However, millisecond accuracy is not sufficient for all applications requiring synchronization. To provide precise time synchronization, the Precision Time Protocol (PTP) was then proposed [1], which operation principle is similar to NTP, but provides new features to meet microsecond synchronization accuracy.

Given the emergence of distributed and real-time applications requiring tight time synchronization performance, there is a growing need to study time synchronization for these applications in a low-cost manner. PTP is currently one of the most investigated synchronization protocols, but there is still a lack of module compliant with the OMNeT++ INET framework to study PTP-based applications over wired and wireless networks. In [2], the authors developed a PTP simulation model over IEEE 802.11 networks. However, the PTP module was added using 802.11 modules specifically without following the OMNeT++ node structures, and the implementation is currently not available to the best of our knowledge. In our implementation, PTP can be used with any communications technology (e.g., Ethernet, 802.11, etc.), since PTP is an application layer protocol and uses the above network layers. In this paper, we present our developed PTP module for OMNeT++ INET, tested with the most recent version (INET 2.6 for OMNeT++ 4.4). To show some illustrative simulation results, we investigate the PTP performance under the presence of variable network loads with (and without) quality-of-service (QoS) support, and propose a new mechanism to improve the PTP accuracy.

The remainder of the paper is structured as follows. In Section II, we overview key background information on the Precision Time Protocol (PTP). In Section III, the PTP implementation model for OMNeT++ is next described. The simulation results are then provided in Section IV. Conclusions are finally drawn in Section V.

This work was supported by NSERC Postdoctoral Fellowship No. 453711-2014.

Corresponding author: Martin Lévesque, School of Information Sciences, University of Pittsburgh, Pittsburgh, PA 15260 (email: levesque@pitt.edu).

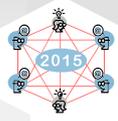

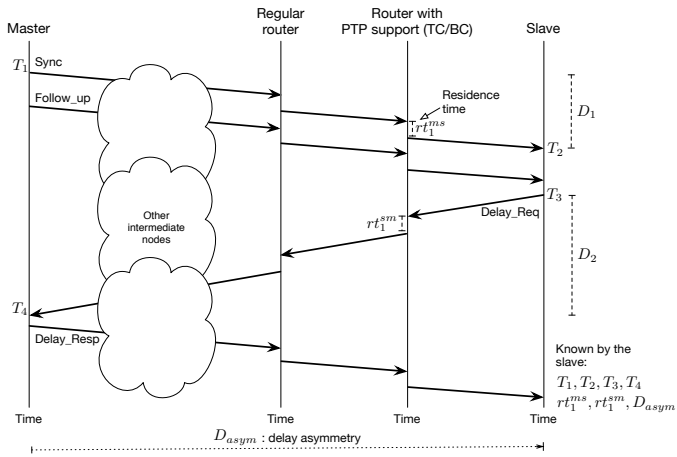

Fig. 1: The Precision Time Protocol (PTP) over a network with a subset of the routers having PTP support. (TC: Transparent clock, BC: Boundary clock)

## II. Background

In this section, we describe some key PTP notions, which we investigate in this paper using our developed OMNeT++ module. The Precision Time Protocol (PTP) version 2, IEEE 1588 [1], enables precise synchronization of clocks over heterogeneous systems with accuracy in the microsecond to submicrosecond range. PTP was proposed to meet tighter synchronization accuracy compared to the widely used NTP protocol, where NTP targets distributed applications to meet millisecond synchronization requirements. On the other hand, PTP was designed to reach precise synchronization requirements for industrial automation, power systems, and telecommunications applications. In a fashion similar to NTP, PTP operates at the application layer. Further, each PTP node contains one or many ports, and communicates with other PTP nodes in a given network by implementing a synchronization protocol described hereafter.

### A. PTP Synchronization Protocol

The main procedures of the PTP synchronization protocol, depicted in Fig. 1, are listed as follow:

- The master node periodically transmits a *Sync* message to the slave nodes containing the sent time $T_1$.
- A *Follow_up* message which contains $T_1$ is optionally sent depending on the timestamp processing mechanism.
- When $T_1$ arrives at a given slave node, the received time $T_2$ is recorded.
- The slave node then sends a *Delay_req* message containing $T_3$ to the master clock.
- Finally, the master node records the reception time $T_4$ and sends it back to the slave node in a *Delay_resp* message.

---

[1]The RFCs and relevant information on NTP are available online: http://www.ntp.org/.

Both the NTP and PTP protocols follow a similar strategy to update the clocks. First, the offset time at a given slave node $k$ is approximated by:

$$\Theta_k = \frac{D_1 - D_2}{2},\qquad(1)$$

where $D_1 = T_2 - T_1$ and $D_2 = T_4 - T_3$ correspond to the downstream (from the master node to slave node) and upstream (from a slave node to master node) delays, respectively. To correct its clock, a given slave node $k$ adjusts its local time $t_k$ to:

$$t_k \leftarrow t_k - \Theta_k.\qquad(2)$$

It is worth noting that Eq. (1) approximates the offset precisely if the messages experience symmetrical delays, that is, if both $D_1$ and $D_2$ are close. The presence of asymmetrical delays can significantly degrade the synchronization accuracy.

### B. PTP Delay and Offset Improvement Mechanisms

PTP introduces three main improvement mechanisms to mitigate the negative effects of asymmetric links on the synchronization accuracy:

- *Residence time at intermediate nodes:* The residence time at a given router corresponds to the time duration a PTP packet resides in a switching fabric from the input port to the output port. In Fig. 1, two residence times are recorded, $rt_1^{ms}$ and $rt_1^{sm}$, which are measured at the router with PTP support.
- *Asymmetric delay parameter, $D_{asym}$:* If the asymmetric delay properties are known in a given network, an asymmetric delay parameter can be used [3], which corresponds to the *delayAsymmetry* field in IEEE 1588.
- *Peer-to-peer path correction:* Peer-to-peer transparent ports measure the link propagation delays. Such a mechanism helps at reducing asymmetry, at the expense of an increased cost.

All of these techniques can be integrated in Eq. (1), as described in [1].

### C. Performance

One significant metric to quantify the PTP performance is the synchronization error, which is the average time deviation between the slaves and master clocks. The PTP performance can vary significantly depending on the network and conditions (traffic load, asymmetry, timestamping method, etc.). In testbeds following most of the PTP standard recommendations, a synchronization accuracy of approximately 50 ns was achieved [4], [5]. Further, a synchronization precision of 2 ns under ideal conditions was obtained by implementing all PTP features in hardware, including the timestamping function [6]. However, under the presence of multi-hop communications with non PTP routers, significantly worst synchronization accuracies of 450 $\mu s$ were measured under the presence of asymmetrical delays [7]. The investigation of the PTP performance is convenient by using an event-driven simulator

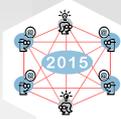

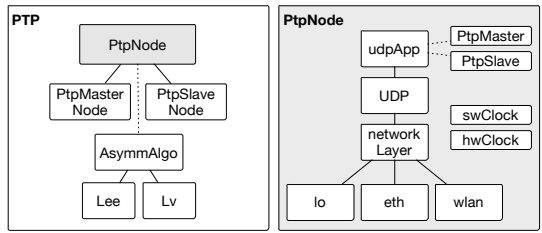

Fig. 2: Main components of the PTP simulation model.

such as OMNeT++, especially to evaluate new mechanisms, as we show shortly in Section IV.

## III. Implementation Details

We next describe our PTP simulation model, ptp++[2], which architecture is depicted in Fig. 2. We adapted and extended the module developed in [8]. The main components of our model are described as follow:

- *PtpNode:* This class models a PTP node, which can be either a master or slave. As PTP operates at the application layer, both the master and slave applications are modeled as a User Datagram Protocol (UDP) application (*udpApp*). In order to be compliant with the OMNeT++ INET 2.6 networking framework, *PtpNode* follows a structure similar to *NodeBase*, thus implementing the overall UDP/Internet Protocol (IP) stack structure, including the application layer (*udpApp*), UDP, network layer, Medium Access Control (MAC), and physical layer.

- *Software and hardware clocks:* As the goal of PTP is to synchronize clocks, a hardware clock needs to be modeled at each node. The implemented hardware clock (*hwClock*) can take different drift distributions based on the model proposed by [8], [9]. The software clock (*swClock*) takes the hardware clock signals as input, and a processing delay can be added to model variable software impairments.

- *AsymmAlgo:* As we mentioned in the previous section, asymmetrical connections can significantly degrade the PTP accuracy. Recently, several mechanisms (e.g., Lee [10] and Lv et al. [11]) have been proposed to detect and mitigate the negative effects of the asymmetry on the PTP accuracy. These mechanisms periodically send extra control messages after the PTP execution. Thus, we developed the simulation model such that currently proposed and new asymmetry mitigation techniques can be evaluated.

Further, a *StatsCollector* module records the clock deviations in order to evaluate the accuracy performance under different network conditions. When a software clock time varies, the synchronization error is computed and added in the *StatsCollector*. In the current implementation, three statistic output files are generated at the end of a given simulation:

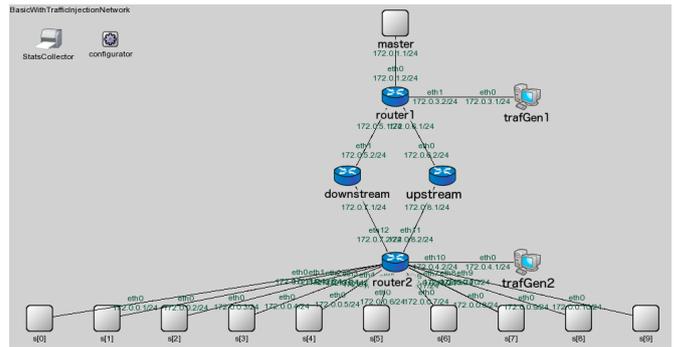

Fig. 3: Simulation scenario consisting of 10 slaves synchronizing to a master node over intermediate Ethernet routers.

- *Summary text file:* The average synchronization error, standard deviation, minimum, and maximum values are appended in a general statistic file. If multiple configurations are executed, such a file can be used for plotting results under variable conditions.

- *Probability distribution function:* Using the recorded synchronization errors, probabilities are computed to represent the distribution of the synchronization errors.

- *Deviation vector:* An OMNeT++ vector of the synchronization errors is also recorded such that the errors can be visualized at the end of a given simulation.

Fig. 3 depicts a sample scenario consisting of 10 slave nodes synchronizing to a master node over three Ethernet routers. Depending on the load variations in the network, the PTP accuracy varies significantly, as we will show in the next section. Further, as recommended in the PTP standard, to improve accuracy, PTP messages must be processed as soon as they arrive in the router queues, where PTP messages are filtered with high quality-of-service (QoS) class priority. We implemented a class-based queue *PtpPrioritizedQueue* which processes PTP messages first by following a deep packet inspection procedure, which we investigate in the following results section.

## IV. Simulation Results

In this section, we investigate the network depicted in Fig. 3 in terms of PTP synchronization accuracy under variable network loads using our developed PTP simulation model. All communications interfaces are based on Ethernet with 100 Mbps full duplex capacity. The communications links are symmetric in terms of propagation and transmission delays. The asymmetry component we consider corresponds to the variable queuing delays experienced mainly in the router switching fabrics (e.g., nodes *router1*, *router2*, *downstream*, and *upstream*). In order to vary the network load, we have two traffic generators, *trafGen1* and *trafGen2*, which exchange messages between each other using the intermediate routers. Further, we configure the network routing tables such that the packets coming to *router1* and destined to nodes *s[1..10]* and *trafGen2* are routed to the *downstream* and *router1* nodes. Similarly, the packets coming in *router2* and destinated upstream are routed to the *upstream* and *router1* nodes. Therefore, the downstream and upstream delays experienced by the PTP messages vary depending on the traffic load generated by the *trafGen1* and *trafGen2* nodes. Further, to simulate realistic and bursty traffic patterns, the traffic generators follow a Pareto distribution such that the average interarrival time between the generated packets equals $\mu = \frac{a \cdot b}{a-1}$, where $a = 1.5$ and $b$ are the shape and scale of the Pareto distribution, respectively.

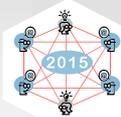

Fig. 4a) depicts the synchronization accuracy under variable upstream and downstream load conditions without using any QoS mechanism[3]. When the traffic load is close to 0 Mbps in the upstream and downstream directions, as expected, the synchronization error is below 10 $\mu$s. However, as the traffic increases, the synchronization error grows, especially when the network is under asymmetrical traffic load conditions (e.g., when the upstream load equals 90 Mbps with 0 Mbps downstream load). To improve these performance, we next configure prioritized queues in order to process PTP messages first, as discussed in the previous section. Fig. 4b) shows significant accuracy improvement while using priority queues, where the synchronization error varies between 0-80 $\mu$s compared to 0-400 $\mu$s while not using any QoS mechanism. As in the previous results (without using any QoS), when the traffic load is asymmetrical, the synchronization error increases. However, we observe that when priority queues are used at high loads, the synchronization error decreases compared to medium loads. This is due to the fact that at high load, the probability that a transmission of a non PTP packet occurs, while a PTP packet arrives at the same time, is higher.

We next propose a new mechanism called *class probing*, which consists of sending an extra non PTP message (and thus low priority) prior to sending a PTP message in order to increase the transmission probability of non PTP message in both directions, to improve asymmetry and consequently the synchronization accuracy. We show a comparison of using this technique vs. using the PTP standard in Fig. 5. We observe a significant improvement of sending an extra non PTP message prior to sending a PTP message, which will be investigated more extensively in future work.

## V. Conclusions

In this paper, we described our PTP simulation model which extends the OMNeT++ / INET framework. Our implementation follows the PTP standard and allows to measure the protocol under different network configurations and conditions. We shown, using the implemented model, that the synchronization accuracy gets significantly affected by variable network loads, and using a prioritized QoS mechanism significantly improves the accuracy. The simulation model allows to investigate new PTP mechanisms under different settings at low-cost. The framework can also be used for time-aware applications, or other protocols requiring tight synchronization. For instance,

---

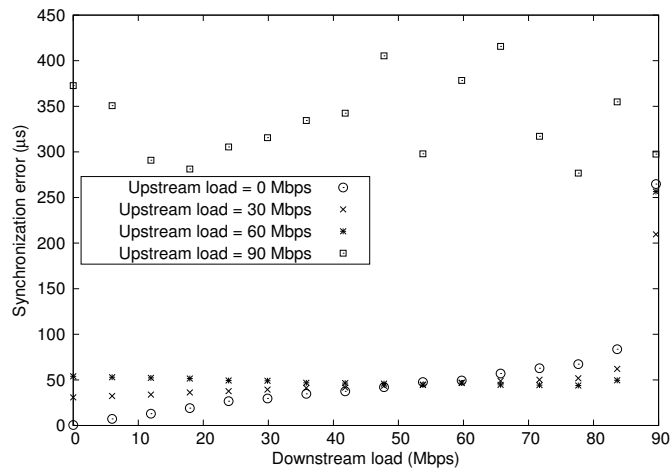

(a) Without QoS.

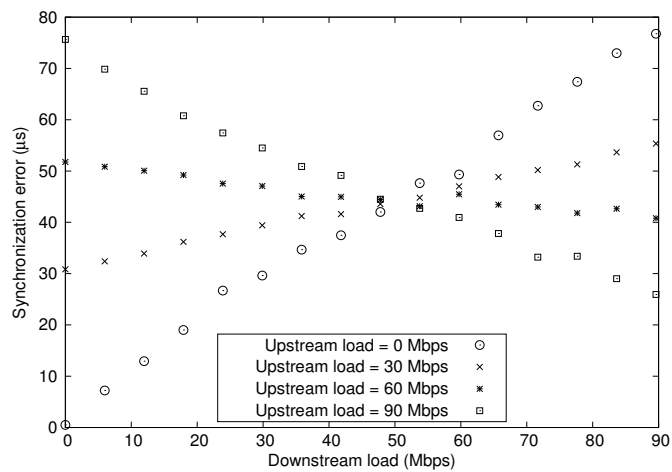

(b) With QoS.

Fig. 4: Synchronization accuracy with variable upstream and downstream network loads.

some security protocols exchange timestamps to detect replay attacks [12]. Further, the emerging Internet-of-Things applications require to timestamp events, thus using a time synchronization protocol such as PTP improves the simulation realism with non perfect synchronized clocks.

---

[3]For each given upstream/downstream pair, we repeated the experiment 15 times with different random seed numbers.

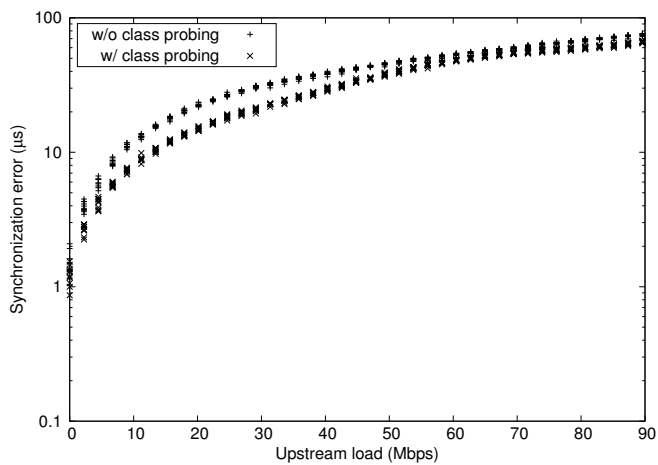

Fig. 5: Improvement of the synchronization accuracy by using a class probing mechanism.